\documentclass[epj]{svjour}

\usepackage[english]{babel}
\usepackage{graphicx}
\usepackage{fancyhdr}
\usepackage{amsmath}
\usepackage{amsfonts}
\def\makeheadbox{{
 }}
\def\mytitle{My title}
\def\myauthors{My name}
\def\mytype{My type of session}
\def\mysession{My session}

\def\mytitle{Impact of non-standard neutrino interactions on future oscillation experiments} 
\def\myauthors{Toshihiko Ota}    
\def\mytype{Contributed Talk}    
\def\mysession{Flavor Physics}

\begin{document}
\title{Impact of non-standard neutrino interactions on future oscillation experiments}
\author{
J. Kopp\inst{1}\thanks{\emph{Email: jkopp@mpi-hd.mpg.de}}
\and
M. Lindner\inst{1}\thanks{\emph{Email: lindner@mpi-hd.mpg.de}}
\and
T. Ota\inst{1}\thanks{\emph{Email: toshi@mpi-hd.mpg.de}} (Speaker)%
\and
J. Sato\inst{2}\thanks{\emph{Email: joe@phy.saitama-u.ac.jp}}
}                     
%
%
\institute{
Max-Planck-Institut f\"ur Kernphysik, Postfach 10 39 80, 69029
Heidelberg, Germany
\and 
Department of Physics, Saitama University, Shimo-Okubo 255,
Sakura-ku, Saitama, 338-8570, Japan
}
%
\date{}
\abstract{
We study the performance of reactor and superbeam neutrino experiments
in the presence of non-standard interactions (NSI). We find that for
some non-standard terms, reactor and superbeam experiments would yield
conflicting result in the $\theta_{13}$ determination, while in other
cases, they may agree well with each other, but the resulting value for 
$\theta_{13}$ could be far from the true value. Throughout our discussion,
we pay special attention to the impact of the complex phases of the NSI
parameters and to the observations at the near detector. 
} 
\maketitle
%
\section{Introduction}
\label{intro}

In lepton flavour physics, one of the most anticipated upcoming results
is the measurement or constraint of the mixing angle $\theta_{13}$
by reactor and accelerator neutrino experiments. Within the context of
standard oscillations, the physics goals of these types of experiments
overlap partly, but if physics beyond the standard model should affect
neutrino oscillations, we may obtain independent information from them
because they observe the different oscillation channels, and new
physics can therefore affect them in differing ways. In this talk,
we will study the question of how non-standard effects could
modify the results of reactor and superbeam experiments, and
will specifically address the following questions:
\begin{itemize}
  \item What will it mean if reactor and accelerator experiments
    yield conflicting results?
  \item If they give consistent results, how rigid is the
    standard oscillation interpretation?
\end{itemize}
In order to approach this subject in a model independent way,
we will introduce effective four Fermi interactions (usually called
non-standard interactions or NSI~\cite{Grossman,original-Epsilon-m}),
which are assumed to be induced by physics beyond the standard model
at some high energy scale.

\begin{figure*}[htb]
\includegraphics[width=0.45\textwidth,angle=0]{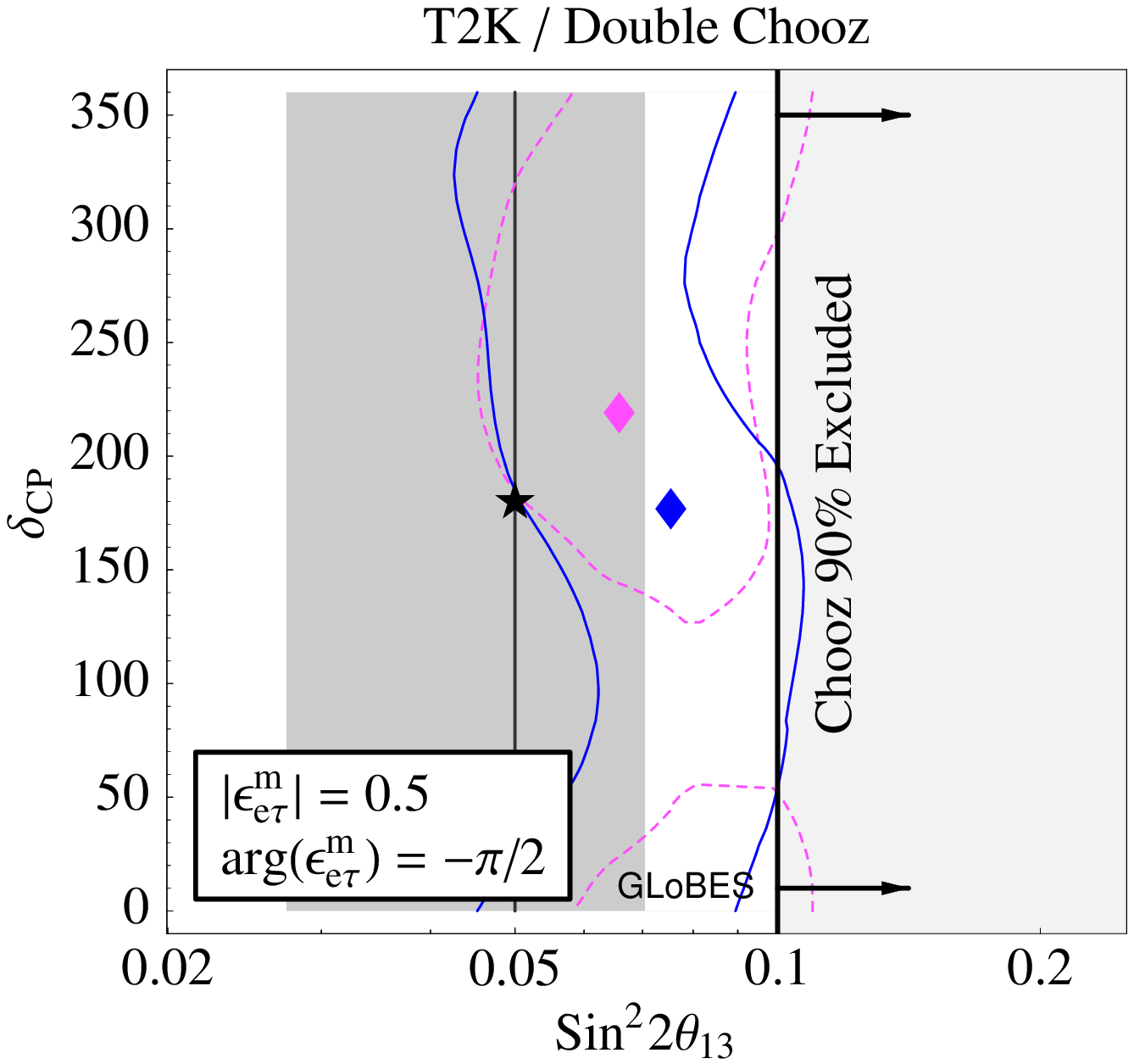}
\hfill
\includegraphics[width=0.45\textwidth,angle=0]{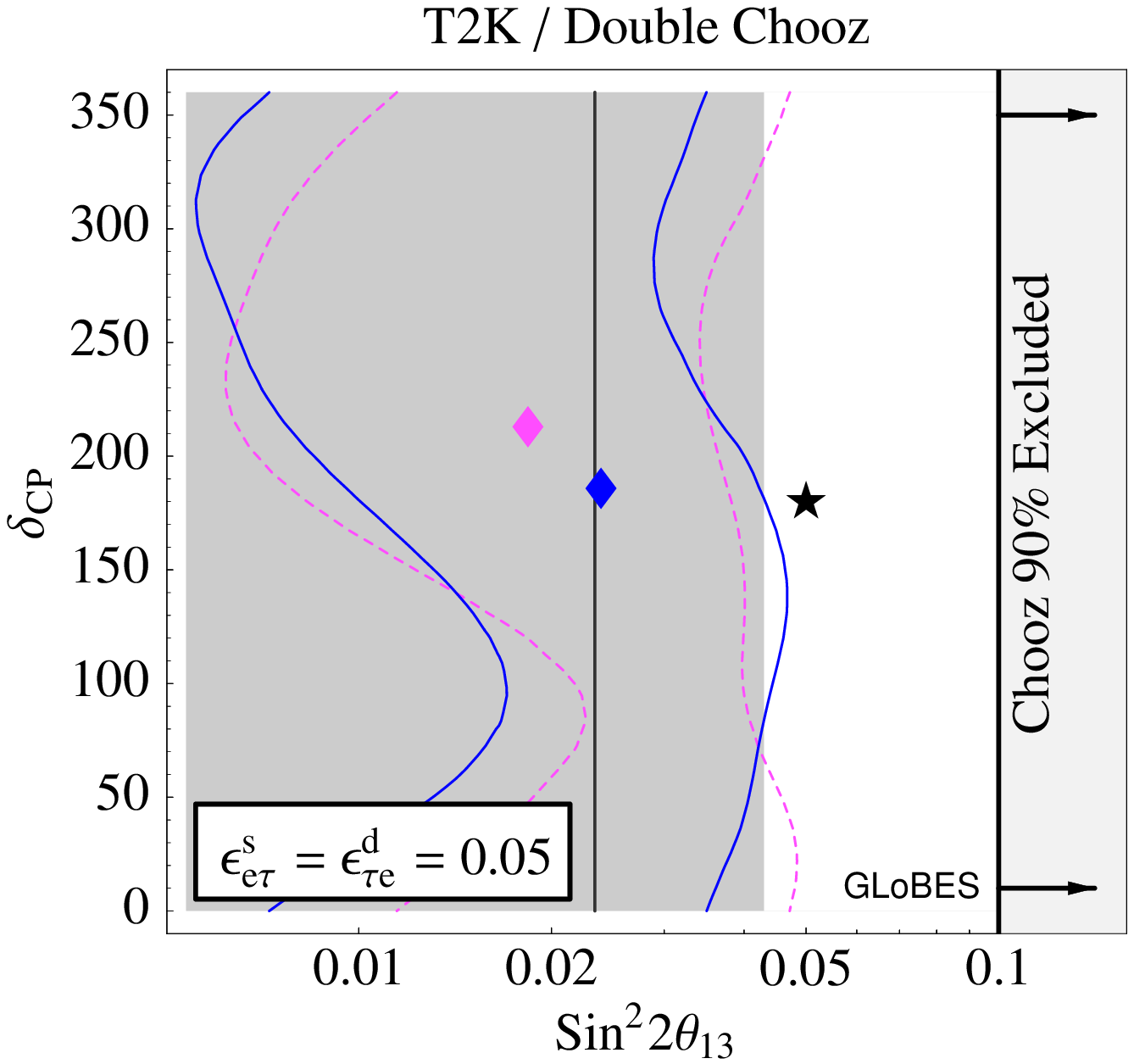}
\caption{Examples for a discrepancy (left) and a common offset (right)
 in the standard oscillation fits for {\sf Double Chooz} and {\sf T2K}.
 In the left panel, corresponding to the NSI parameter $\epsilon^{m}_{e\tau}$,
 the best-fit value of {\sf Double Chooz} (vertical solid line) is
 located far from the best-fit value of {\sf T2K} (coloured diamonds). 
 In the right panel, with $\epsilon^{s}_{e\tau}$ and $\epsilon^{d}_{e\tau}$,
 both fits coincide with each other but have a common offset from the 
 true value (denoted by the black star).}
\label{Fig:offset-discrepancy}
\end{figure*}

For NSI mediated by charged currents, the neutrino states involved in the
production and detection processes will not be pure flavour states, but
can be described as mixed-flavour states, which we call $|\nu^{s} \rangle$
for the {\it source state} and $\langle \nu^{d} |$ for the {\it detection
state}. They are defined as~\cite{Grossman}
\begin{align}
|\nu^{s}_{\alpha} \rangle
 =
|\nu_{\alpha} \rangle 
 + 
 \sum_{\gamma=e,\mu,\tau} 
 \epsilon^{s}_{\alpha \gamma}
 | \nu_{\gamma} \rangle, 
\label{eq:def-epsilon-s}
\\
\langle
 \nu_{\alpha}^{d} |
 =
 \langle \nu_{\alpha}|
 +
 \sum_{\gamma=e,\mu,\tau}
 \epsilon^{d}_{\gamma \alpha}
 \langle \nu_{\gamma}|.
\label{eq:def-epsilon-d}
\end{align}
Here, the index $\alpha$ stands for the flavour of the accompanying
charged lepton in the respective process.

We can also introduce NSI mediated by neutral currents, which would
create an additional, possibly flavour-violating matter potential
of the form~\cite{original-Epsilon-m}
\begin{align}
(V_{\rm NSI})_{\beta \alpha}
 =
 \sqrt{2} G_{F} N_{e} \epsilon^{m}_{\beta \alpha}.
\label{eq:def-epsilon-m}
\end{align}

The $\epsilon$ parameters represent the ratio between the NSI amplitude
and the standard amplitude. For example, $\epsilon^{s}_{\alpha\beta}$ and
$\epsilon^{d}_{\beta \alpha}$ are the degrees of ``contamination''
of the source and detection states with ``wrong'' flavours, and 
$\epsilon^{m}_{\beta \alpha}$ denotes the ratio between the
non-standard and standard matter effects in the propagation
Hamiltonian. Taking into account the NSI shown in
Eqs.~\eqref{eq:def-epsilon-s} -- \eqref{eq:def-epsilon-m},
we can calculate the oscillation probability 
for $\nu_{\alpha}^s \rightarrow \nu_{\beta}^d$ as
\begin{align}
P_{\nu_{\alpha} \rightarrow \nu_{\beta}}
 =
 \left|
 \langle \nu^{d}_{\beta}|
 {\rm e}^{-{\rm i} (H_{\rm SO} + V_{\rm NSI}) L}
 | \nu^{s}_{\alpha} \rangle
 \right|^{2}.
\end{align}
Here, $H_{\rm SO}$ is the standard neutrino propagation Hamiltonian,
which is parameterized by the standard oscillation parameters.
Although the Lorentz structure of the effective NSI operators
is not fixed by experimental constraints, we will only consider
$(V-A)(V-A)$ type interactions in this talk.\footnote{The case of 
NSIs with different Lorentz structures is discussed in Ref.~\cite{NSI}.}
Under this assumption, we can treat the $\epsilon$ parameters
as energy-independent. Moreover, the following
relation between the NSI in the neutrino beam source and
those in the detection process holds:
\begin{align}
\epsilon^{s}_{\alpha \beta}
=
(\epsilon^{d}_{\beta \alpha})^{*}.
\end{align}
We will use this relation as a constraint in the numerical simulations
which we are going to present in the next section.

\section{NSI-induced offsets and discrepancies in $\theta_{13}$ fits}
\label{sec:1} 

\begin{figure*}[htb]
\includegraphics[width=0.36\textwidth,angle=0]{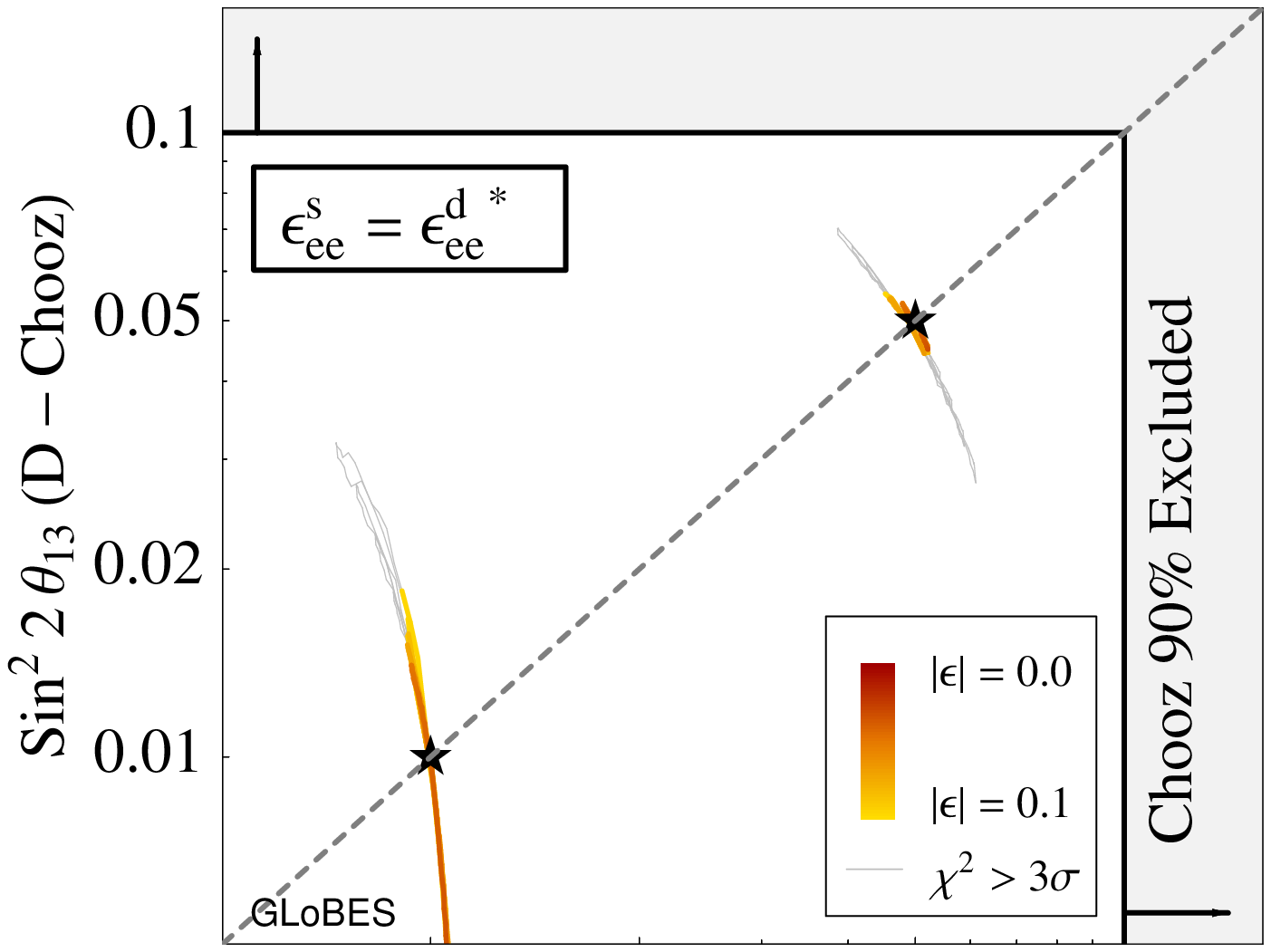}
\hspace{-0.8cm}
\includegraphics[width=0.36\textwidth,angle=0]{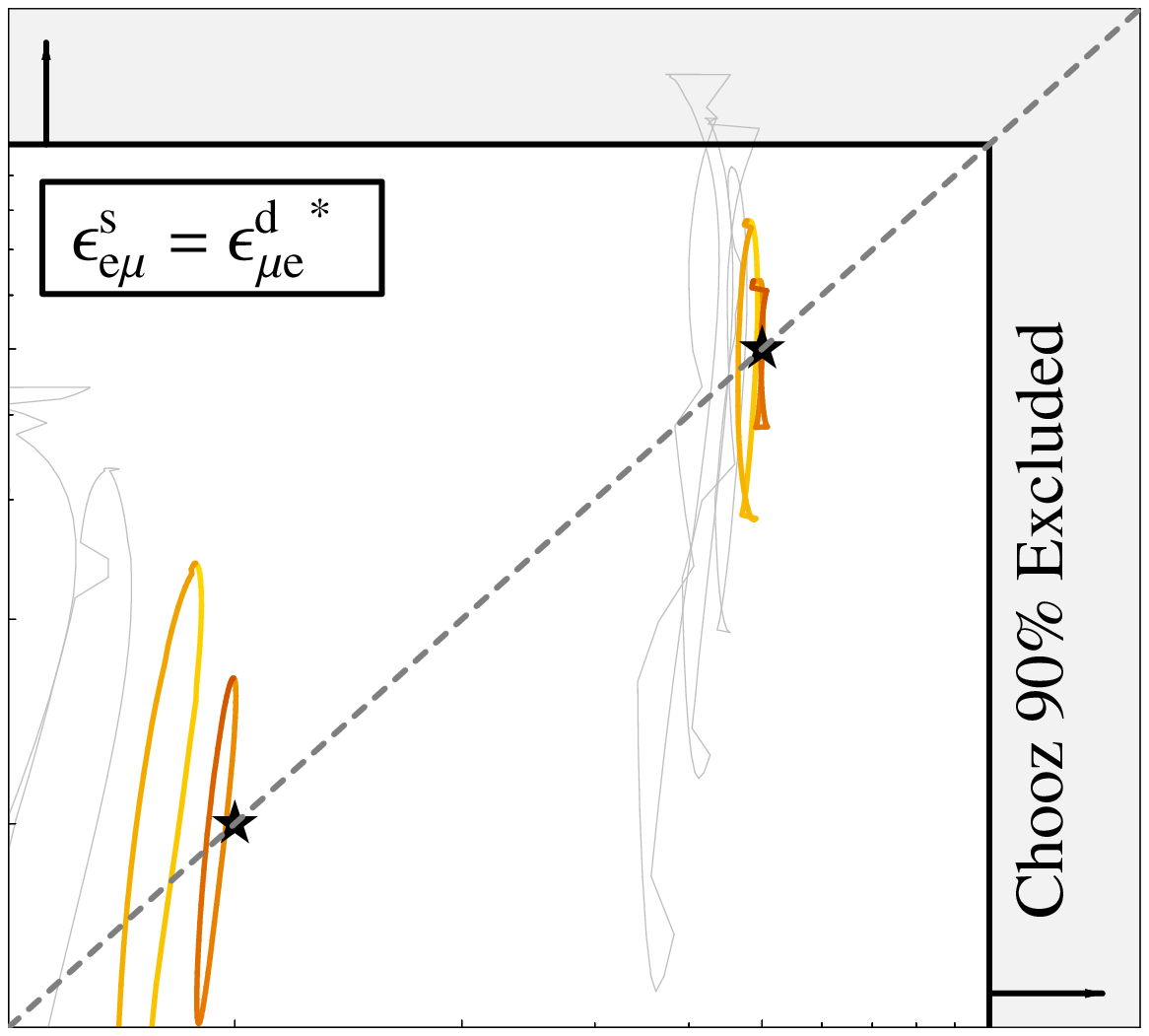}
\hspace{-0.8cm}
\includegraphics[width=0.36\textwidth,angle=0]{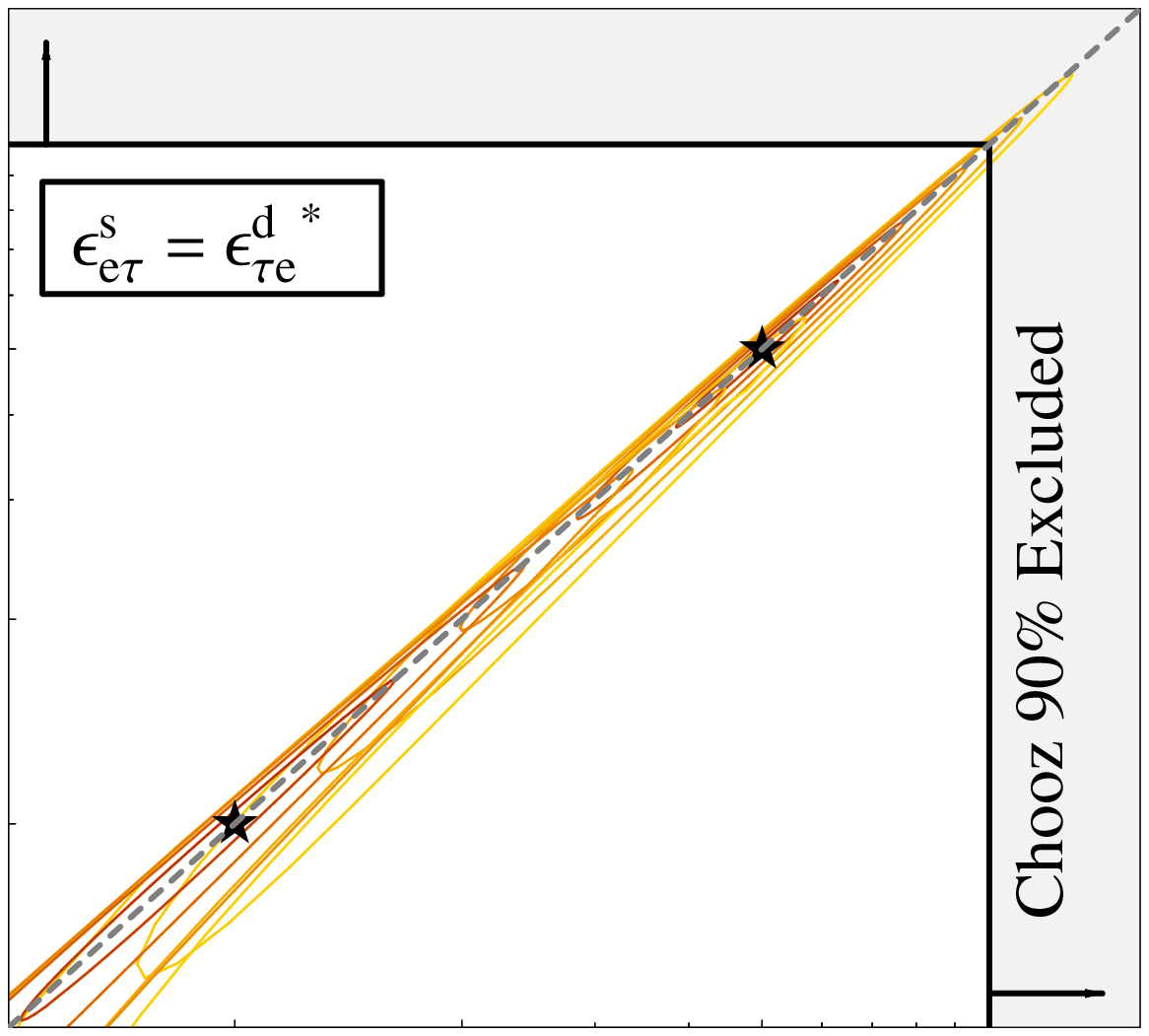}
\vspace{-1.1cm}\\
\includegraphics[width=0.36\textwidth,angle=0]{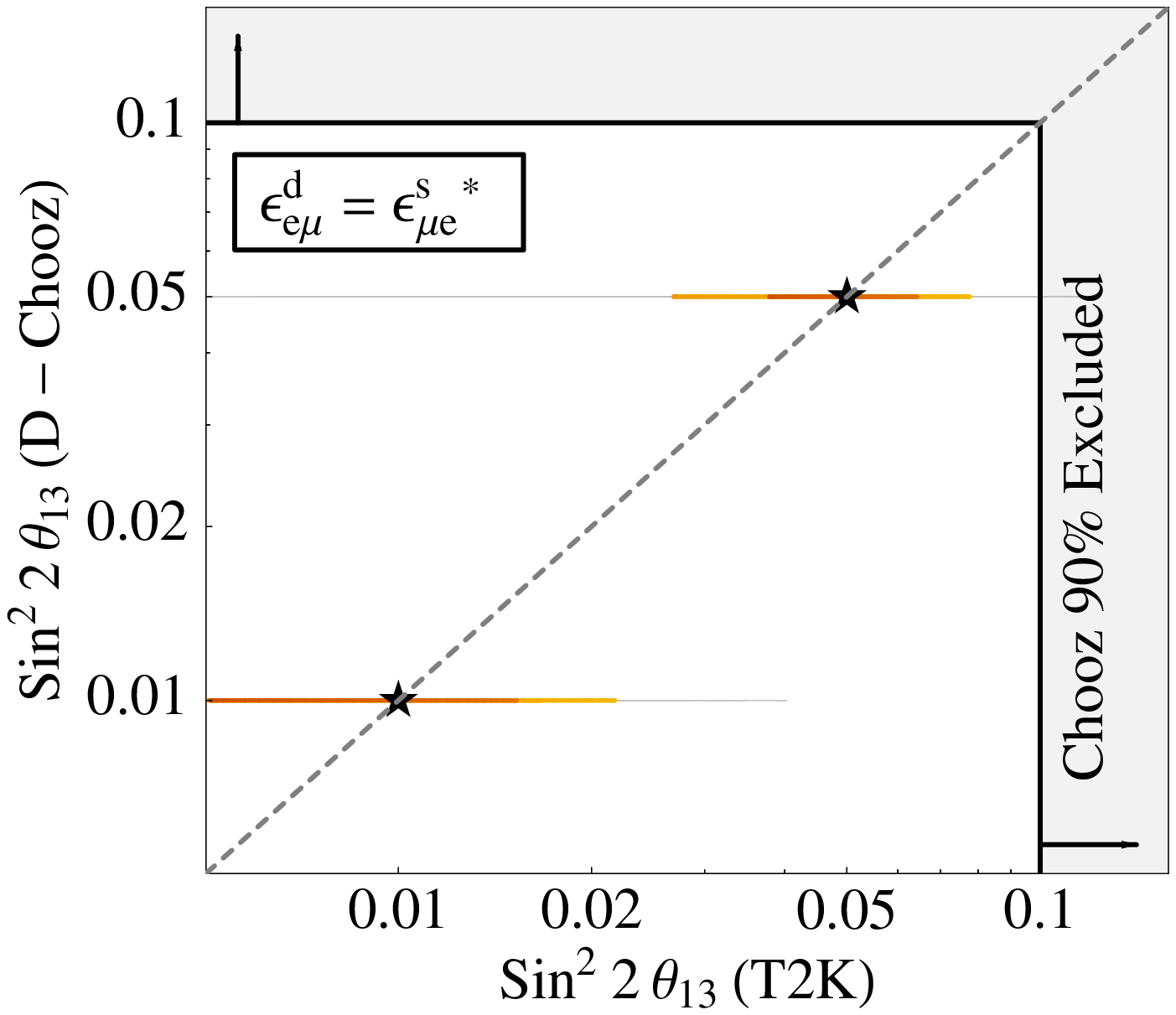}
\hspace{-0.8cm}
\includegraphics[width=0.36\textwidth,angle=0]{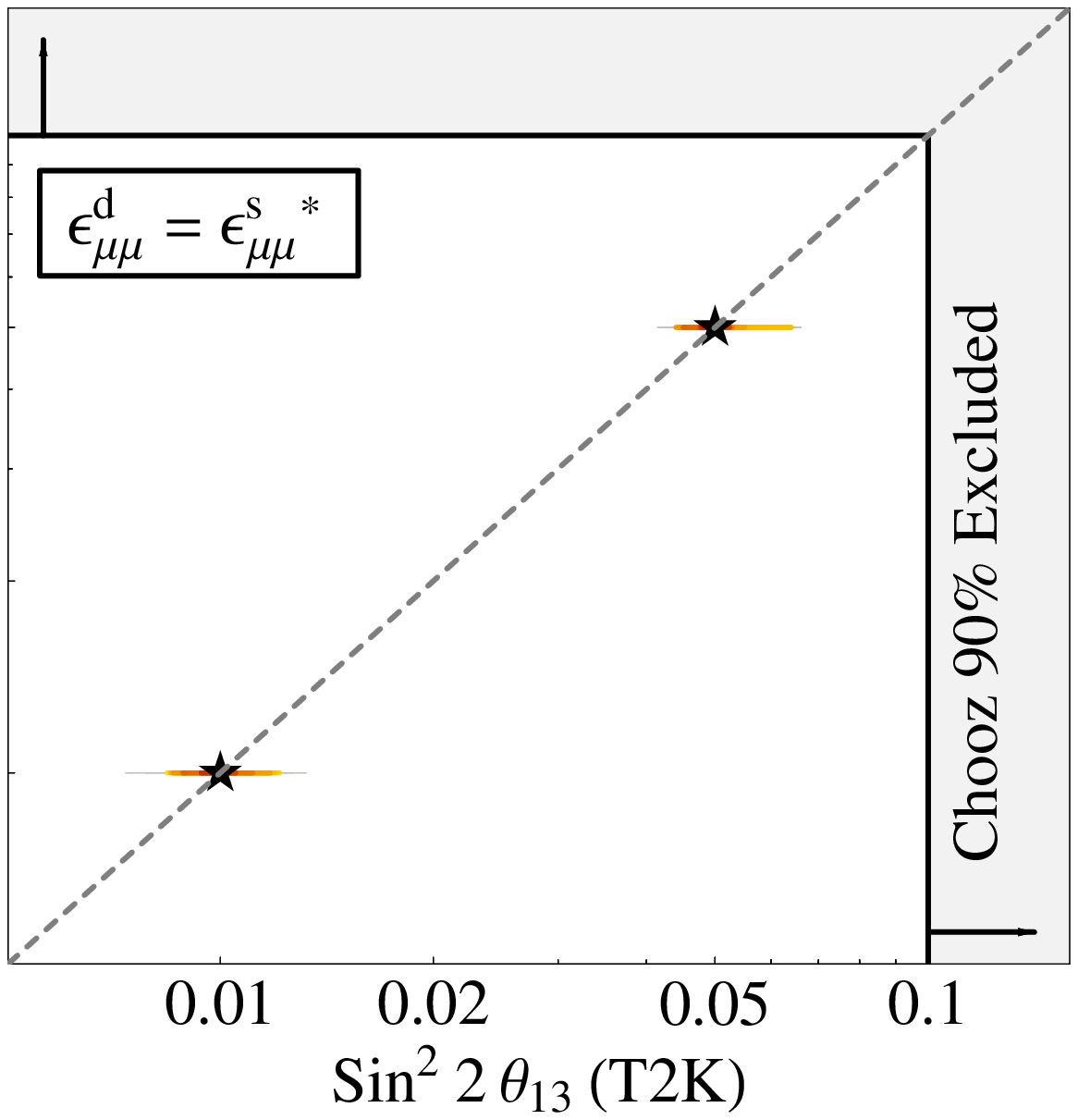}
\hspace{-0.8cm}
\includegraphics[width=0.36\textwidth,angle=0]{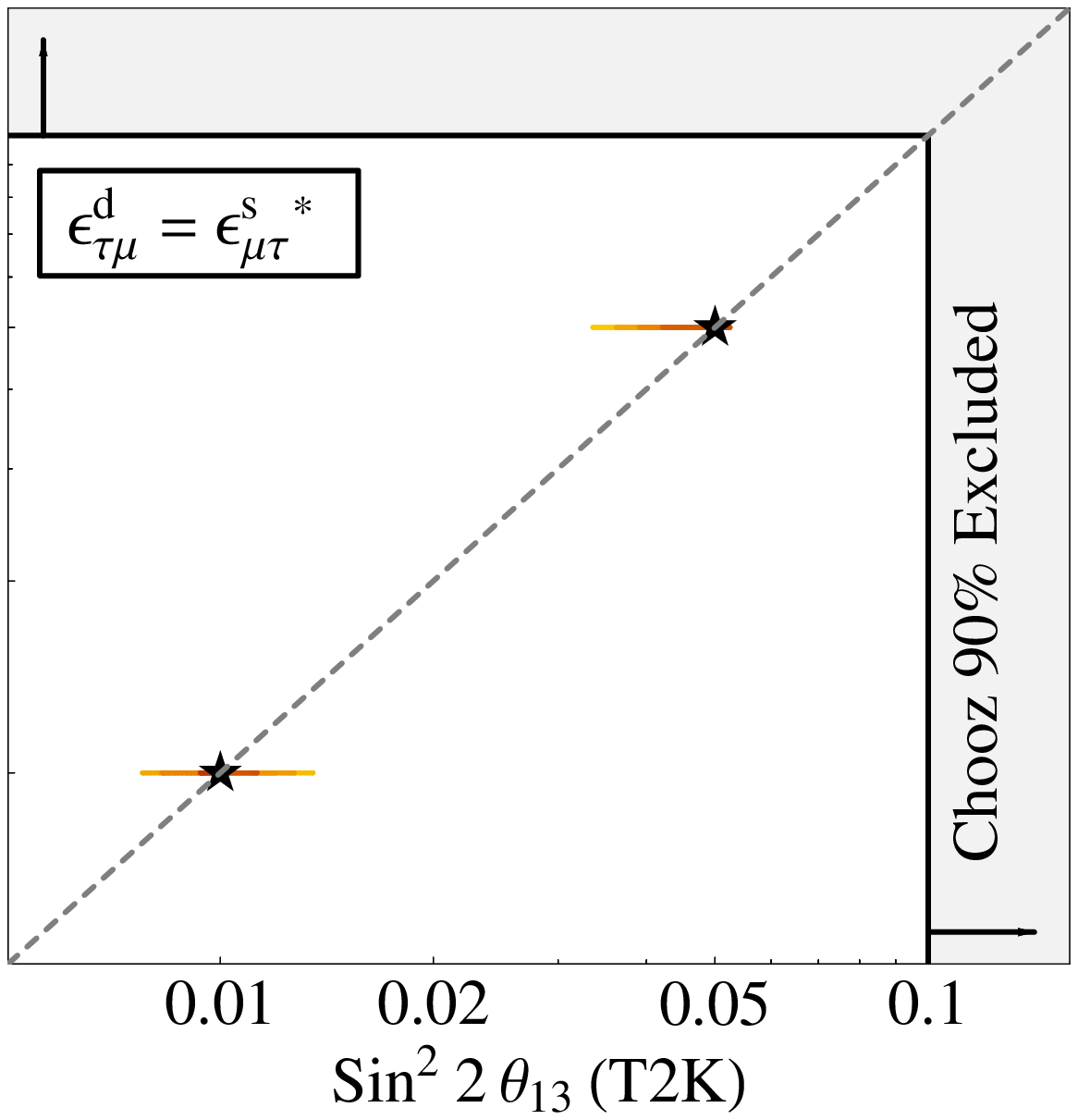}
\caption{Distribution of the best-fit values for $\theta_{13}$ in
  {\sf T2K} and {\sf Double Chooz} in the presence of the NSI parameters
  $\epsilon^{s}_{\alpha\beta}$ and $(\epsilon^{d}_{\beta \alpha})^{*}$.
  The black stars indicate the assumed true values of $\sin^{2} 2\theta_{13}$
  which are taken to be 0.05 and 0.01.}
\label{Fig:best-fit}
\end{figure*}
To fully assess the high-level consequences of 
NSI for realistic reactor and superbeam experiments, we have
performed numerical simulations using the {\sf GLoBES} 
software~\cite{GLoBES}. In this talk, we will concentrate on only
two experiments, namely {\sf T2K} and {\sf Double Chooz}.
To analyse the simulated data, we define the following $\chi^{2}$ function
\begin{align}
\chi^{2} =& \min_{\lambda}
 \sum_{j}^{\text{channel}}
 \sum_{i}^{\text{bin}}
 \frac{
 |
 N_{ij} (\lambda^{\text{true}}, \epsilon^{\text{true}})
 - 
 N_{ij} (\lambda, \epsilon=0)|^{2}}
{
 N_{ij}(\lambda^{\text{true}}, \epsilon^{\text{true}})} \nonumber \\
&+
{\rm Priors},
\end{align}
where $N_{ij}$ denotes the number of neutrino 
events in the $i$-th energy bin for oscillation channel $j$,
the vector 
$\lambda = (\theta_{12}, \theta_{13}, \theta_{23},
\delta_{\rm CP}, \Delta m_{21}^{2}, \Delta m_{31}^{2}, \vec{b})$
contains the six standard oscillation parameters and the systematical biases
$\vec{b}$, and $\epsilon$ represents the non-standard parameters.
For the following plots, $\chi^{2}$ has been marginalised over all standard
oscillation parameters (except, of course, those for which we are interested
in confidence regions rather than best-fit points) and over all systematical
biases. We keep the NSI parameters fixed at 0 in the fit because we want to
study the impact of NSI on a {\it standard oscillation} fit. Unless
indicated otherwise, we calculate the simulated event rates using
the following reference values ({\it true values}) for the standard
oscillation parameters~\cite{best-fit},
\begin{align}
\sin^{2} 2\theta_{12}^{\text{true}} = 0.84,  &\qquad
\sin^{2} 2 \theta_{23}^{\text{true}} = 1.0,  \nonumber\\
\sin^{2} 2 \theta_{13}^{\text{true}} = 0.05, &\qquad
\delta_{\rm CP}^{\text{true}} = 0,           \nonumber\\
(\Delta m_{21}^{2})^{\text{true}} &= 7.9 \cdot 10^{-5} \text{ eV}^{2}, \nonumber \\
\left| (\Delta m_{31}^{2})^{\text{true}} \right| &= 2.5 \cdot 10^{-3} \text{ eV}^{2}.
\end{align}
We assume the true hierarchy to be normal. Although we keep
$\delta_{\rm CP}^{\rm true}$ fixed, this does not limit the generality
of our results, because the oscillation probabilities typically depend
only on combinations of $\delta_{\rm CP}$ and the phases of the NSI
parameters. Thus, a variation of $\delta_{\rm CP}$ has the same effect
as a variation of $\mathrm{arg}(\epsilon)$.

In our simulation of {\sf T2K}, the {\sf Super-Kamiokande} far detector
and a 1.0 kton water Cerenkov near detector are simulated separately.
We introduce a common 10\% uncertainty on the neutrino flux and a common
20\% error on the number of background events in the $\nu_{e}$ appearance
channel. Also for {\sf Double Chooz}, we simulate the near and far detectors
separately. We introduce a 2.8\% correlated flux normalization error
as well as uncorrelated 0.6\% fiducial mass errors, and a bin-to-bin
uncertainty of 0.5\%. More details on the input parameters of our
simulations are given in Ref.~\cite{NSI}.

Two exemplary outcomes of the simulation  are plotted in
Fig.~\ref{Fig:offset-discrepancy}. In both panels, we have chosen
the ``true'' parameter values $\sin^{2} 2 \theta_{13} = 0.05$ and
$\delta_{\rm CP} = \pi$, as indicated by the black star.
The vertical black line and the coloured diamonds represent
the best-fit values for {\sf Double Chooz} and {\sf T2K} ,
respectively, while the shaded vertical band and the coloured contours
are the corresponding 90\% confidence regions. Blue symbols
and lines stand for the {\sf T2K} normal hierarchy fit, and dashed
magenta ones are for an inverted hierarchy fit. The NSI contributions
are assumed to be $\epsilon^{m}_{e\tau} = 0.5 {\rm e}^{-{\rm i} \pi/2}$
in the left hand plot and $\epsilon^{s}_{e\tau} = \epsilon^{d}_{\tau e}
= 0.05$ in the right hand plot. In the first case, the best-fit value
of {\sf Double Chooz} conincides with the ``true'' value because
reactor experiments are sensitive neither to standard nor to
non-standard matter effects. On the other hand, the $\theta_{13}$ fit
of {\sf T2K} deviates significantly, indicating that $\epsilon^{m}_{e\tau}$
has a large impact on this experiment, even though {\it standard} matter
effects are only sub-dominant due to the relatively short
baseline.\footnote{Standard and non-standard matter effects become more
significant at baselines of $\mathcal{O}(10^{3})$~km, which are, for
example, considered for a neutrino factory. For recent studies of NSI
in this context, see e.g.~Refs.~\cite{NSI-nuFACT,Ribeiro2007}.}
The discrepancy is so large, that the best-fit value of {\sf T2K} is
almost ruled out at 90\% confidence level by the reactor measurement. 
In the case shown on the right hand side of Fig.~\ref{Fig:offset-discrepancy},
both experiments agree well with each other, but their fits suffer from a
common NSI-induced offset, which even leads to an erroneous exclusion
of the true $\theta_{13}$.

\begin{figure}[htb]
\includegraphics[width=0.45\textwidth,angle=0]{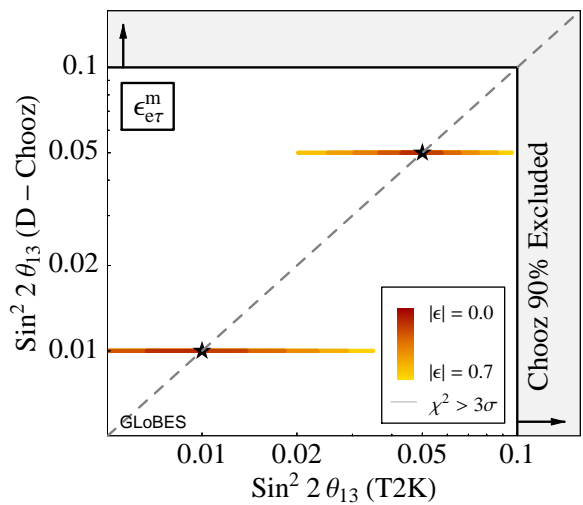}
\caption{Distribution of the best-fit values for $\theta_{13}$ in
  {\sf T2K} and {\sf Double Chooz} in the presence of the NSI parameter
  $\epsilon^{m}_{e\tau}$.}
\label{Fig:best-fit-epsilon-m}
\end{figure}

Let us now proceed to a more systematic analysis, which we present in 
Figs.~\ref{Fig:best-fit} and \ref{Fig:best-fit-epsilon-m}.
In these figures, we plot the distribution of the best-fit $\theta_{13}$
values of both experiments for different values of the NSI parameters.
Each curve corresponds to fixed $|\epsilon|$, and
varying $\mathrm{arg}(\epsilon)$, with dark red curves denoting
$|\epsilon| \sim 0$, and yellow curves corresponding to $|\epsilon|$
equal to its current upper bound, which is 0.1 for
$\epsilon^s_{\alpha\beta}$~\cite{bound1} and 0.7 for
$\epsilon^m_{e\tau}$~\cite{bound}. If the $\chi^{2}$ function
for one of the experiments exceeds $3\sigma$, the colour of the
curves has been changed to grey, regardless of $|\epsilon|$. This
indicates that in these cases, a $3\sigma$ {\it discovery} of
the respective non-standard effect is actually possible. The
black stars indicate the chosen ``true'' $\theta_{13}$.
Let us briefly comment on the different cases shown in
Figs.~\ref{Fig:best-fit} and \ref{Fig:best-fit-epsilon-m}.
If $\epsilon^{s}_{ee}$ and $\epsilon^{d}_{ee}$ are introduced
(upper left panel of Fig.~\ref{Fig:best-fit}), 
they modify the $\bar{\nu}_e$ flux in {\sf Double Chooz}, and
lead to a wrong measurement of the $\nu_e$ contamination
in the {\sf T2K} beam by the near detector. Therefore, {\sf T2K}
is affected indirectly. The correct treatment of the near
detectors is crucial here to fully assess the impact of NSI.
Similarly, the effect of $\epsilon^{s}_{e\mu}$ and
$\epsilon^{s}_{e\tau}$ can only be fully understood if the
near detectors are taken into account~\cite{NSI}.
The upper right panel of Fig.~\ref{Fig:best-fit} shows
that $\epsilon^{s}_{e\tau}$ and $\epsilon^{d}_{\tau e}$
are particularly dangerous because they lead to a common
offset in the reactor and in the superbeam  (the best
fit values are distributed along the diagonal). We have already
seen an example of this effect in Fig.~\ref{Fig:offset-discrepancy}.
The NSI parameters $\epsilon^{s}_{\mu \alpha}$ and $\epsilon^{d}_{\alpha \mu}$
affect only {\sf T2K}, hence the distribution of the corresponding
$\theta_{13}$ fits is constrained to horizontal lines in our plots.
Among the non-standard matter effects, only $\epsilon^{m}_{e\tau}$
can give a significant effect, because the other parameters are
either subdominant in the oscillation probability, or are already
well-constrained experimentally. The effect of $\epsilon^{m}_{e\tau}$
is a possibly large shift of the $\theta_{13}$ fit in {\sf T2K}, as
shown in Fig.~\ref{Fig:best-fit-epsilon-m}. 

\begin{figure}
\includegraphics[width=0.45\textwidth,angle=0]{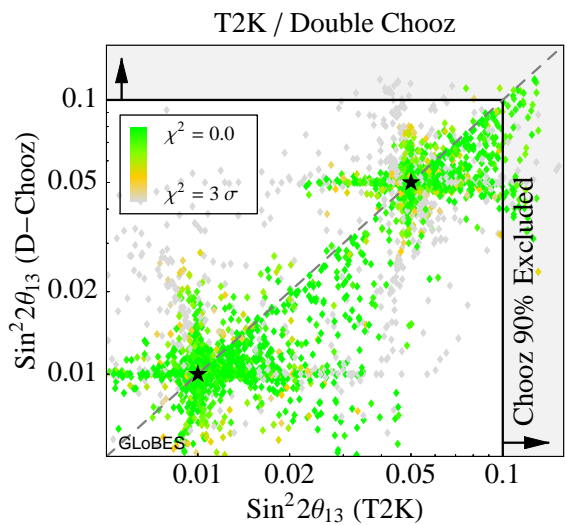}
\caption{Scatter plot of the $\theta_{13}$ best-fit values of
 {\sf T2K} and {\sf Double Chooz} in the presence of random
 combinations of NSI parameters. It is obvious that NSI can induce
 ostensible discrepancies between the two experiments (off-diagonal points),
 or common offsets (close-to-diagonal points) which would lead
 to consistent, but wrong results.}
\label{Fig:scatter-plot}
\end{figure}

So far, we have only considered situations in which one NSI
parameter is dominant, and all the others are negligible. In contrast
to this, Fig.~\ref{Fig:scatter-plot} shows the distribution of the 
$\theta_{13}$ fits in situations where many NSI parameters are
present simultaneously. For each point, the moduli
$|\epsilon^{s,d,m}_{\alpha\beta}|$ were chosen randomly on a logarithmic
scale between $10^{-8}$ and the respective experimental upper
bounds~\cite{bound,bound1}, while the phases are linearly distributed
between 0 and $2\pi$. The colour coding indicates the $\chi^{2}$ value
for each point. We can see from the plot that, for part
of the parameter space, the NSI effect can actually be discovered,
i.e.~the standard oscillation interpretation does not give a good fit.
However, there are also a lot of points with a low $\chi^2$, and among
these are many which exhibit a clear discrepancy between the
$\theta_{13}$ fits of {\sf Double Chooz} and {\sf T2K},
and others which correspond to a common offset of the fit values.

\section{Conclusion}

In this talk, we have discussed the impact of non-standard neutrino
interactions on reactor and accelerator experiments, in particular on
{\sf Double Chooz} and {\sf T2K}. We have introduced flavour violating
four-fermion interactions with $(V-A)(V-A)$ type Lorentz structure,
and considered the impact of these NSI operators on the neutrino
production, propagation, and detection stages. We have performed
detailed numerical simulations, taking
into account parameter correlations, degeneracies, systematic errors,
and a realistic treatment of the near detectors.
We have found that NSI can have a sizeable impact on future experiments
if the corresponding couplings are close to their current upper bounds,
and if the  complex phases do not conspire to cancel them.
There are scenarios in which a clear discrepancy between 
{\sf Double Chooz} and {\sf T2K} arises, but we have also found
situations in which both experiments would agree very well,
but the derived $\theta_{13}$ has a significant offset from
the true value. To detect this kind of problems, a third experiment,
complementary to the other two, would be required.
Our discussion shows that reactor and superbeam measurements,
which might seem to be redundant in the standard oscillation framework,
turn out to be highly complementary once non-standard effects are 
considered.

More details on the impact of non-standard interactions on reactor
and superbeam experiments can be found in Ref.~\cite{NSI}, in which
we also discuss the discovery potential for NSI, paying
specific attention to its dependence on the complex phases. Moreover,
Ref.~\cite{NSI}~provides an intuitive understanding of the impact of the
different $\epsilon$ parameters, and contains analytic formulae to
substantiate the numerical results.

%

\end{document}